# Automated Network Service Scaling in NFV: Concepts, Mechanisms and Scaling Workflow

O. Adamuz-Hinojosa[1,2], J. Ordonez-Lucena[1,2], P. Ameigeiras[1,2], J.J. Ramos-Munoz[1,2], D. Lopez[3], J. Folgueira[4].

[1]*Research Center on Information and Communication Technologies, University of Granada.*
[2]*Department of Signal Theory, Telematics, and Communications, University of Granada.*
[3]*Technology Exploration & Standards, Telefónica I+D-Global CTO.*
[4]*Transport & IP Networks, Telefónica I+D-Global CTO.*

*Abstract* - Next-generation systems are anticipated to be digital platforms supporting innovative services with rapidly changing traffic patterns. To cope with this dynamicity in a cost-efficient manner, operators need advanced service management capabilities such as those provided by NFV. NFV enables operators to scale network services with higher granularity and agility than today. For this end, automation is key. In search of this automation, the European Telecommunications Standards Institute (ETSI) has defined a reference NFV framework that make use of model-driven templates called Network Service Descriptors (NSDs) to operate network services through their lifecycle. For the scaling operation, an NSD defines a discrete set of instantiation levels among which a network service instance can be resized throughout its lifecycle. Thus, the design of these levels is key for ensuring an effective scaling. In this article, we provide an overview of the automation of the network service scaling operation in NFV, addressing the options and boundaries introduced by ETSI normative specifications. We start by providing a description of the NSD structure, focusing on how instantiation levels are constructed. For illustrative purposes, we propose an NSD for a representative NS. This NSD includes different instantiation levels that enable different ways to automatically scale this NS. Then, we show the different scaling procedures the NFV framework has available, and how it may automate their triggering. Finally, we propose an ETSI-compliant workflow to describe in detail a representative scaling procedure. This workflow clarifies the interactions and information exchanges between the functional blocks in the NFV framework when performing the scaling operation.

# 1. Introduction

Network softwarization is an unprecedented techno-economic transformation trend that takes advantage of commodity hardware, programmability and reusability of software to provide cost optimizations and service innovation in next-generation networks. Network Functions Virtualization (NFV) is a key enabler in this trend. It brings novel practices for flexible and agile Network Service (NS) provisioning and management.

The European Telecommunications Standards Institute (ETSI) has defined a reference NFV architectural framework [1] for the purpose of management and orchestration of NSs in multi-vendor, multi-network environments. This framework consists of three Management and Orchestration (MANO) functional blocks that make use of model-driven templates to deploy and operate multiple instances of different NSs (and their constituents) over a common infrastructure. The NS Descriptor (NSD) is the name of the template used for NSs. With the information gathered in a NSD, the MANO blocks are able to manage NS instances throughout their lifecycle with great agility and full automation.

Scaling is a key lifecycle management operation in NFV. Scaling with NFV allows operators to automatically resize NSs at runtime to handle load surges with performance guarantees. This brings dynamicity and cost reduction compared to today's scaling practices, where NS capacity is statically over-dimensioned for the highest predictable traffic peak. To achieve the required automation when scaling, an appropriate model for the NSD is needed.

On one hand, most of works dealing with scaling focus on mechanisms/strategies for virtual resource estimation (e.g., [2]) and allocation (e.g., [3]). The policy-based rules and input data that these proposals use for the scaling operation are not retrieved from an NSD; instead, they are specified manually for every NS. This may led to less agile and more error-prone scaling solutions, where the automation in NFV is not fully exploited.

On the other hand, there are no existing works that analyze the effect the NSD modeling has in the NS scaling operation. The existing MANO platforms based on the NFV framework (e.g., Open Baton, OSM, ONAP or Tacker) use their own data modeling languages (e.g., TOSCA, YANG) for their NSDs. This leads to non-compatible workflows for the NS scaling operation, avoiding reusability and portability of scaling solutions across different NFV platforms [4]. To enable their interoperability, ETSI works on the development of normative specifications for the NFV information model, including interface description, and a platform- and technology-agnostic model for the NSD. Understanding this standardized model, and adapting the existing data models to it, is key for successful scaling operations with the existing MANO platforms. In this line, ETSI NFV has recently started in [5] a work targeted at mapping the TOSCA data model with the NSD information model.

The contribution of this paper is twofold. First, we provide an overview of the structure of the ETSI NSD. We address those NSD fields most relevant for scaling, placing emphasis on the instantiation levels. These levels specify the different sizes an NS instance can adopt throughout its lifecycle. This limits the scaling of an NS instance to one of the discrete set of levels defined in the NSD. Thus, their correct design is key to ensure appropriate scaling operations with NFV. To facilitate their understanding, and show how they are constructed in a NSD, we propose a simple example of an NSD. Secondly, we provide an overview of the automated NS scaling operation, analyzing in depth the options and boundaries introduced by the ETSI NSD information model. We show the different scaling procedures that the NFV framework has available, and how they can be triggered in an automated manner. This includes the proposal of an ETSI-compliant workflow for a representative scaling procedure. This workflow clarifies how the different MANO blocks interact, specifying the information they exchange in each step.

This article is organized as follows. First, we present a background of those NFV concepts relevant for scaling. Then, we provide an insight into the NSD. Next, we detail the most relevant NS scaling procedures, and propose a workflow for one of them. Finally, we remark some conclusions. Table 1 lists the acronyms used throughout the paper.

| Acronym | Meaning |
|---|---|
| BSS | Business Support System |
| DRPA | Dynamic Resource Provisioning Algorithm |
| EM | Element Manager |
| ETSI | European Telecommunications Standards Institute |
| MANO | Management and Orchestration |
| NFV | Network Functions Virtualization |
| NFVI | Network Functions Virtualization Infrastructure |
| NFVI-PoP | Network Functions Virtualization Infrastructure Point of Presence |
| NFVO | Network Functions Virtualization Orchestrator |
| NMS | Network Management System |
| NS | Network Service |
| NS-IL | Network Service Instantiation Level |
| NSD | Network Service Descriptor |
| OSS | Operations Support Systems |
| VCD | Virtual Compute Descriptor |
| VDU | Virtual Deployment Unit |
| VIM | Virtual Infrastructure Manager |
| VL | Virtual Link |
| VLD | Virtual Link Descriptor |
| VNF | Virtualized Network Function |
| VNF-IL | Virtualized Network Function Instantiation Level |
| VNFC | Virtualized Network Function Component |
| VNFD | Virtualized Network Function Descriptor |
| VNFFG | Virtual Network Function Forwarding Graph |
| VNFFGD | Virtual Network Function Forwarding Graph |

|        | Descriptor                        |
|--------|-----------------------------------|
| VNFM   | Virtual Network Function Manager  |
| VSD    | Virtual Storage Descriptor        |

Table 1. List of Acronyms

## 2. Background on key NFV concepts

In this section, we describe the concept of NS and the NFV architectural framework.

### 2.1 The Concept of NS

An NS is a composition of network functions. According to ETSI NFV, network functions may be implemented as Virtualized Network Functions (VNFs) and physical network functions. For simplicity, we only consider the former in this paper.

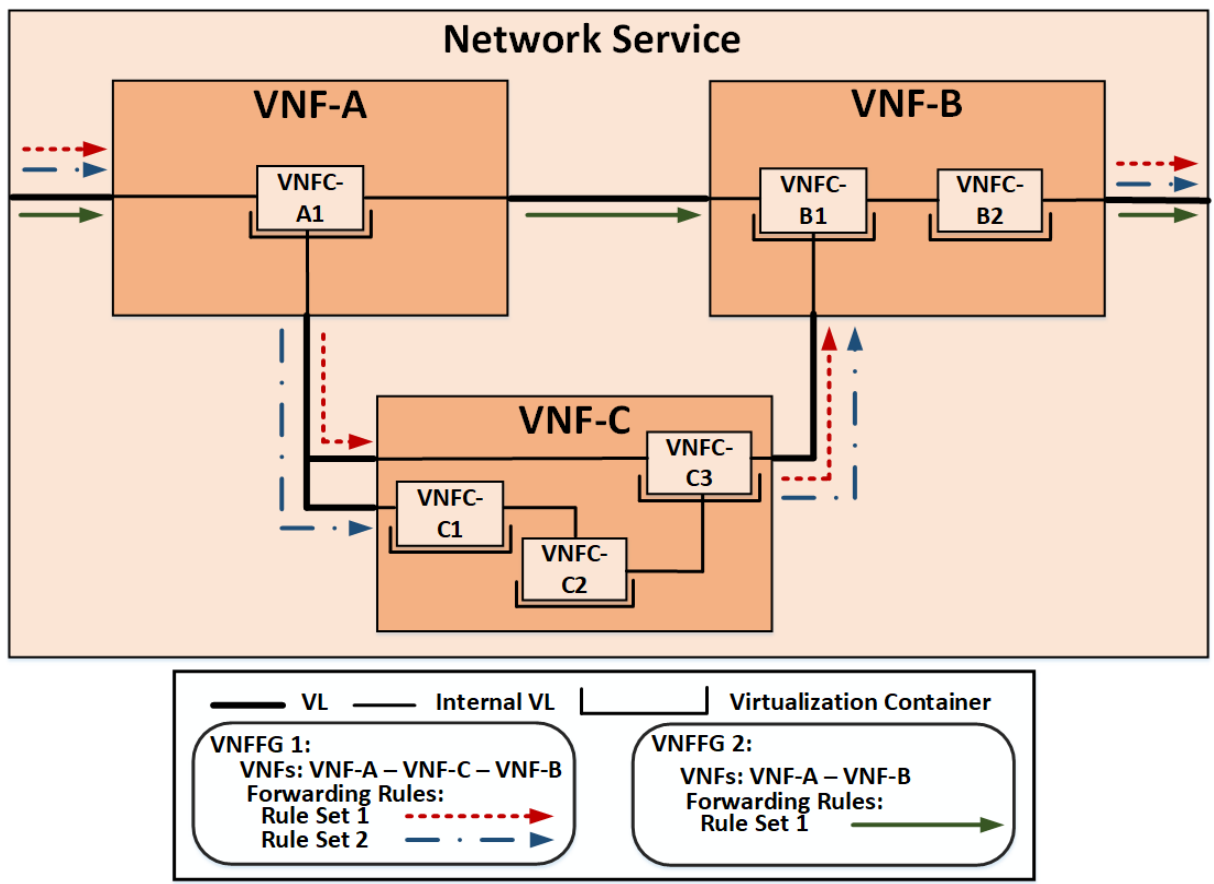

Figure 1. NS internal composition. In this example, we have defined two VNFFGs, and we have associated each with a different network plane: VNFFG1 for user plane traffic, and VNFFG2 for management plane traffic. Note that VNFFG1 includes two set of forwarding rules for traffic steering, enabling the definition of two user plane traffic flows, e.g. for distinct processing.

An NS consists of a set of VNFs, Virtual Links (VLs), and VNF Forwarding Graphs (VNFFGs). VLs are abstractions of physical links that logically connect together VNFs. To specify how these connections are made along the entire NS, one or more VNFFGs are used. A VNFFG describes the topology of (the entire or part of) the NS, and optionally includes forwarding rules to describe how traffic shall flow between the VNFs defined in this topology.

For a fine-grained control of its scalability, performance and reliability, a VNF might be decomposed into one or more VNF Components (VNFCs) [6], each performing a well-defined subset of the entire VNF functionality. Each VNFC is hosted in a single virtualization container, and connected with other VNFCs through internal VLs.

Figure 1 shows an example of an NS with its constituents.

## 2.2 NFV Architectural Framework

ETSI has defined a reference NFV architectural framework [1] for the deployment and operation of NSs. As seen in Fig. 2, this framework consists of three main working domains: the NFV Infrastructure (NFVI) and VNFs, MANO, and Network Management Systems (NMS).

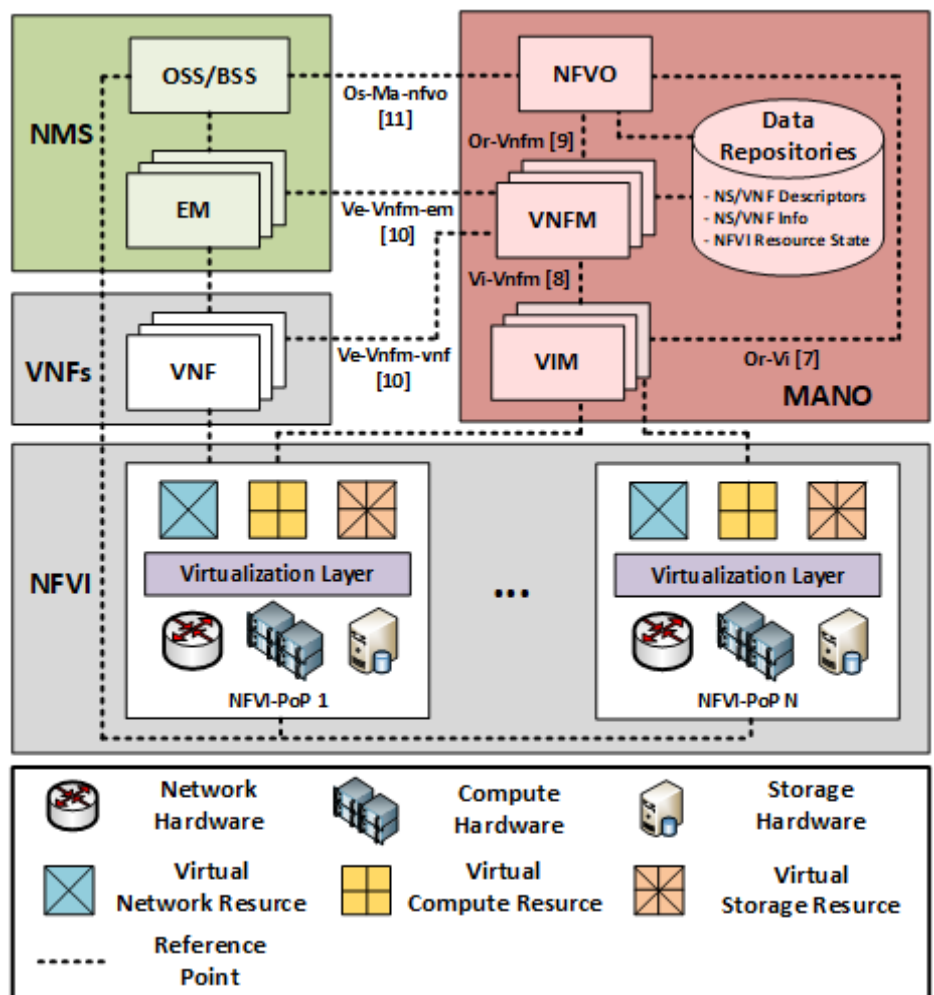

Figure 2. ETSI NFV Architectural Framework. The three working domains, and their constituent functional blocks communicate together using a set of reference points. A detailed description of those reference points targeted for standardization in ETSI NFV can be found in the citations depicted in the figure.

The NFVI is the collection of resources that make up the cloud on top of which VNFs run. With the help of a virtualization layer, the underlying physical resources are abstracted and logically partitioned into virtual resources, used for hosting and connecting VNFs. NFVI might span across several geographically remote NFVI Point of Presences (NFVI-PoPs), enabling multi-site VNF deployments.

MANO focuses on the virtualization-specific deployment and operation tasks in the NFV framework [12]. MANO consists of three functional blocks, including:
- Virtualized Infrastructure Manager(s) (VIM), each managing the resources of one or more NFVI-PoPs.
- VNF Manager(s) (VNFM), focused on the lifecycle management of the VNFs, and responsible for their performance and fault management at virtualized resource level.
- NFV Orchestrator (NFVO), that orchestrates NFVI resources across VIMs, and performs NS lifecycle management.

The MANO also includes data repositories to assist these blocks with their tasks. These repositories include: (a) NS and VNF Descriptors, (b) information about all the NS/VNF instances during their lifecycle (NS/VNF Info), and (c) updated information about the state (allocated/reserved/available) of NFVI resources.

Finally, the NMS focuses on traditional (non virtualized-related) management tasks, orthogonal to those defined in MANO. NMS comprise:
- Element Manager(s) (EM), responsible for the fault, performance, configuration, accounting, and security management of the VNFs at application level.
- Operations/Business Support Systems (OSS/BSS), comprising traditional systems and management applications that help operators to provision and operate their NSs.

# 3. NS Description

In this section, we first study the NSD structure. Then, we show an example of an NSD. For illustrative purposes, we propose an NSD for the NS shown in Fig. 1.

## 3.1 NSD Overview

An NSD is a deployment template that contains machine-processable information used by MANO blocks to create instances of an NS, and operate them throughout their lifetime. An NSD is constructed from a set of attributes and other descriptors (see Fig. 3).

The attributes that an NSD includes enable specifying how to deploy and operate instances of an NS. In this work, we consider those that are most relevant for NS scaling (see Fig. 3):
- *Monitored Info*: specifies the information to be tracked for NS performance and fault management. This information includes resource-related performance metrics (at NS/VNF level), and VNF indicators from NS's constituent VNFs.

- *Auto Scaling Rules*: contain rules that enables triggering scaling actions on an NS instance when a condition involving *Monitored Info* is not satisfied. The NFV information model allows expressing these rules as customized scripts provided at instantiation time. The language used for these scripts shall support conditions involving not only logical/comparison operators with scalar values, but also complex analytical functions able to process statistical data correlated from different sources.
- *Deployment Flavors*: describe specific deployment configurations for the NS. For a more detailed description, see next subsection.

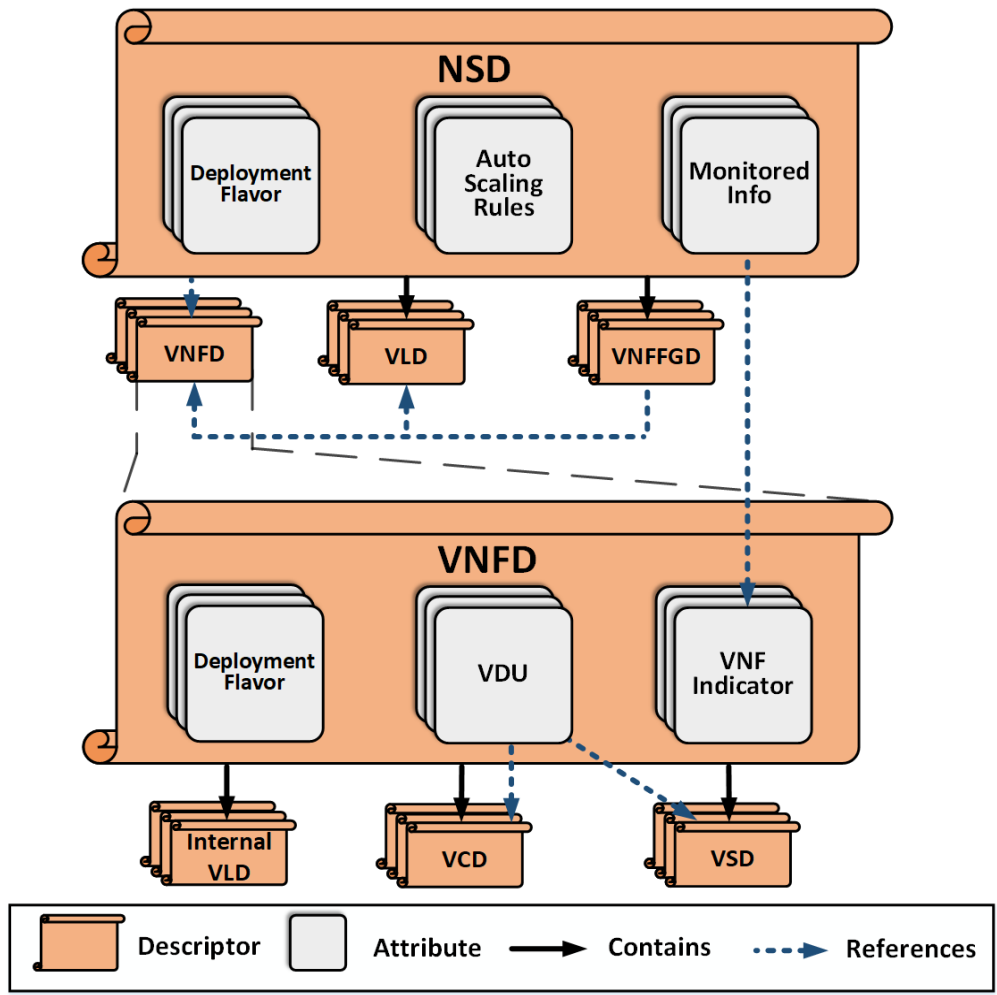

Figure 3. NSD structure. Only the descriptors and attributes that are most relevant for NS scaling are shown.

To describe the deployment and operational behavior of NS constituents, NSD contains and references a set of descriptors, including VNF Descriptors (VNFDs), VL Descriptors (VLDs), and VNFFG Descriptors (VNFFGDs) [13]. A VNFD contains information required to deploy and operate instances of a VNF. A VLD provides information of a VL, including the deployment configurations available for VL instantiation. These configurations are specified through *Deployment Flavors*. Different configurations results in different performance and reliability levels for VNF connectivity. Finally, a VNFFGD references the VNFDs and VLDs for topology description.

From the above descriptors, we concentrate on the VNFDs. The NSD references information of VNFDs that is essential for NS scaling. Similar to an NSD, a VNFD also includes descriptors and attributes.

The descriptors that VNFD contains provide a detailed view on the VNF internal composition. Particularly, a VNFD includes Virtual Compute Descriptors (VCDs), Virtual Storage Descriptors (VSDs), and internal VLDs. The first two specify the virtual compute and storage resources that are needed for VNFC hosting, while the latter specifies the performance requirements for VNFC connectivity.

In terms of attributes, a VNFD includes one or more:
- *VNF Indicators*: represent performance/fault-related events that provide information of the VNF at application level.
- *Virtualization Deployment Units (VDUs)*: describe how to create and operate instances of VNFCs; hence, an VDU can be seen as a VNFC descriptor. A VDU specifies the compute resources (and optionally storage resources) that a virtualization container needs to host a VNFC. To that end, it references one VCD (and optionally one or more VSDs).
- *Deployment Flavors*: similar to those defined in the NSD, but applied to VNFs.

As seen, the models for the NSD and VNFD are very similar: VNFs/VDUs connected by VLDs, and defining various Deployment Flavors. Thus, some initiatives like the Superfluidity project [4] have suggested the idea of having a common, reusable information model for both of them, so that they can share the same interfaces and lifecycle management operations. This model shall be recursive and scalable. ETSI NFV already enables this recursivity and scalability at NS level, with the concept of composite NSs (i.e., an NS composed of smaller, nested NSs) [13]. However, ETSI NFV considers the necessity to maintain different models for NSDs and VNFDs, due to some technical issues that can be found in [13-14].

From the perspective of NS scaling, the Deployment Flavors within an NSD are key attributes, as they contain the instantiation levels permitted for an NS instance. These levels are constructed with information included in the flavors of the VNFDs and VLDs. In the next subsection, we describe the different flavors, studying how they enable the definition of different instantiation levels in the NSD.

### 3.2 Deployment Flavors and Instantiation Levels

As seen earlier, there are three types of deployment flavors: VL flavors, VNF flavors, and NS flavors.

Selecting a VL flavor enables selecting specific QoS parameters (latency, jitter, etc.) and transport reliability for a VL.

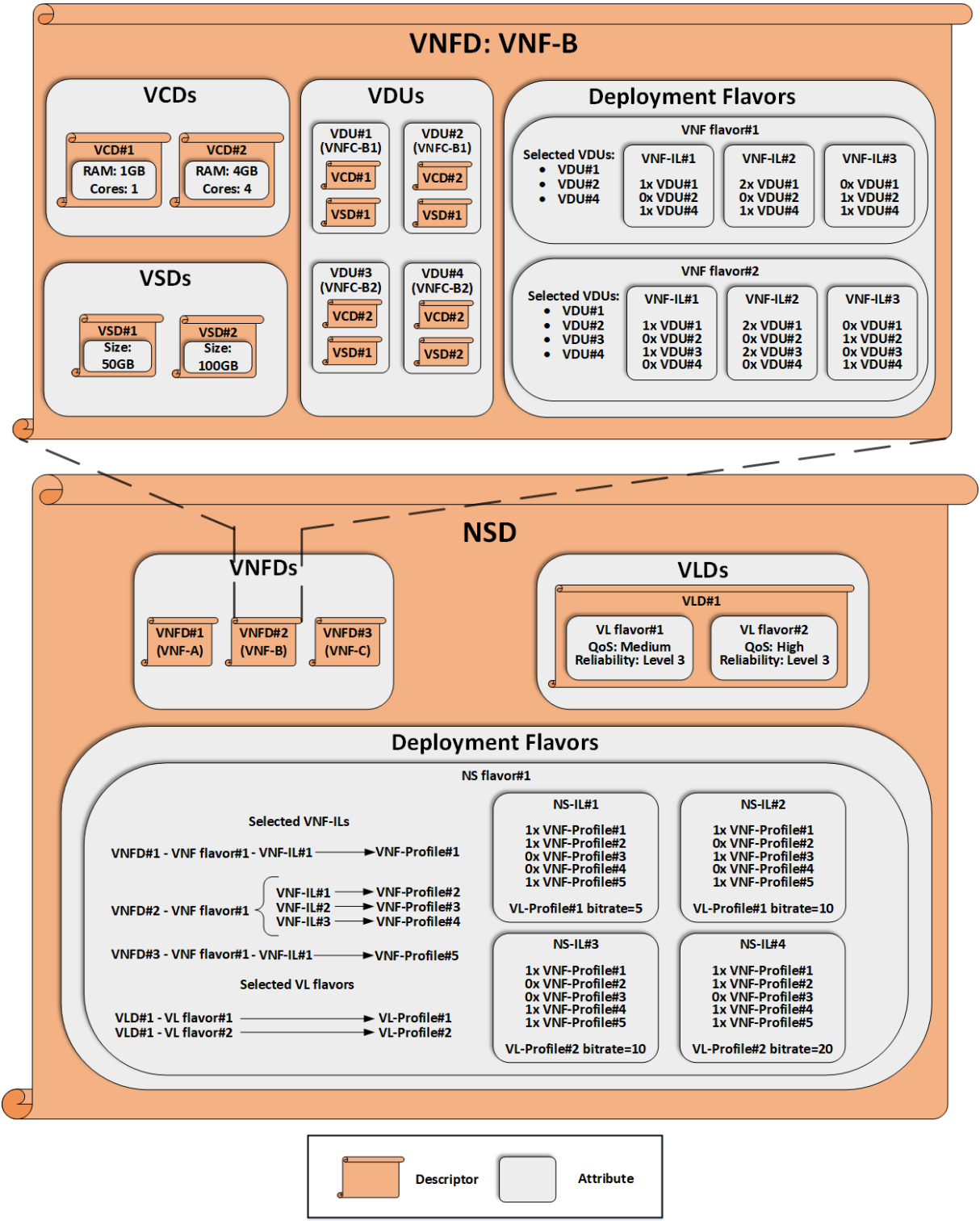

Figure 4. A NSD proposal for the NS given in Fig. 1. Please note that only the most relevant attributes for scaling are shown. For better understandability, the VNF-ILs selected for the NS flavor are referred to as VNF-Profiles. Similarly, the VL flavors selected for the NS flavor are referred to as VL-Profiles. The profile term is also used in ETSI NFV. See [13] for more information.

Each VNF flavor within a VNFD can be used to define a different deployment configuration for a VNF. A given deployment configuration specifies the functionality and the performance level(s) allowed for instantiating the VNF. To specify the VNF functionality (i.e. which features need to be activated for the VNF), the VNF flavor indicates which (subset of)

VNFCs need to be deployed. Particularly, the flavor references the VDUs to be used for their instantiation. To specify the performance level(s) permitted for VNF instantiation (i.e. which amount of resources are needed for each selected VNFC), the flavor defines one or more instantiation levels (VNF-ILs). Each VNF-IL indicates, for each VDU referenced in the flavor, the number of VNFC instances that need to be deployed from this VDU.

Finally, NS flavoring enables adjusting the functionality and the level of performance of an NS. A NS flavor selects the VNFs and VLs to be deployed as part of the NS, and their actual flavors. For each selected VNF flavor, the subset of VNF-ILs to be used is also specified. With this information, an NS flavor defines one or more instantiation levels (NS-ILs). NS-ILs are similar to VNF-ILs, but at NS level. Each NS-IL specifies:
- The number of VNF instances to be deployed from each VNF-IL specified in the NS flavor.
- The VLs required for VNF connectivity, their bitrate, and the VL flavors used for their instantiation.

As seen above, NS-ILs are key for NS scaling. The fact that an NS instance needs to be scaled means that the current NS-IL is no longer valid for that instance, and hence a new NS-IL must be used. In this case, the NFVO must select, among the finite set of NS-ILs predefined in the NS flavor, the optimum one for scaling the instance (see next section). Note that the operator's policy adopted for NS flavor design (the number of NS-ILs, and the differences among them in terms of resource requirements) has a great impact on the NS scaling operation at operation time.

To clarify these concepts, in Fig. 4 we provide an example of an NSD for the NS shown in Fig. 1. For simplicity, we only address the VNFD corresponding to VNF-B: VNFD#2. The rest of VNFDs could be constructed similarly.

VNFD#2 includes four VDUs. From these VDUs, instances of VNFC-B1 and VNFC-B2 with different capacity can be created. For example, VDU#1 and VDU#2 are used to create instances of VNFC-B1 with low/high capacity, respectively. VNFD#2 also includes two flavors: VNF Flavor#1, enabling only VNFC-B1 scaling; and VNF Flavor#2, support the scaling of both VNFC-B1 and VNFC-B2. Each flavor includes three VNF-ILs, differing in their resource requirements.

The NSD references the descriptors of the NS's constituent VNFs: VNFD#1, VNFD#2 and VNFD#3. For simplicity, we assume VNFD#1 and VNFD#3 have each a single flavor with a single VNF-IL. VNF-A, VNF-B, and VNF-C are interconnected through VLs that can be instantiated from the two defined VL flavors. In this example, we consider a single NS flavor that only allows the scaling of VNF-B, restricting VNF-A and VNF-C to a single instance each. This flavor presents four predefined NS-ILs. These NS-ILs enable two scaling cases: (a) increasing/decreasing the capacity of a VNF-B instance, and (b) adding/removing a VNF-B instance. The first three NS-ILs are used for (a), where there is one instance of VNF-B. When NS-IL#3 is reached, the VNF-B instance can no longer increment its capacity. At this point, the only way to scale this NS is by adding a new VNF-B instance, as stated in NS-IL#4. With this NS-IL, the NS instance has two instances of VNF-B.

# 4. NS Scaling Automation

In this section, we describe how the NS scaling operation may be automated with MANO, considering the boundaries that ETSI NFV specifications impose. We detail the input information that the NFVO takes to determine if an NS instance needs to be scaled. Assuming the scaling is required, we show the different scaling procedures that NFVO may trigger. Finally, we propose a detailed workflow to describe one of them, illustrating the messages the MANO blocks exchange in that procedure.

## 4.1 Boundaries and Procedures

Although NS scaling can be manually triggered, automation enables operators to fully exploit NFV benefits. To automate the scaling triggering, NFVO has a customizable software module (e.g. supporting NS-specific code) that runs a Dynamic Resource Provisioning Algorithm (DRPA). The DRPA determines when an operative NS instance needs to be scaled, and the NS-IL which optimizes the scaling of that instance according to a set of criteria. This optimum NS-IL will then be used by the NFVO to trigger an appropriate scaling procedure.

The DRPA takes the following input parameters:
- Performance and fault data, as specified in the Monitored Info attribute (see Fig. 3). This includes periodical resource-related performance metrics [6-9], and/or asynchronous alarms (performance metric-based threshold crossing, and VNF indicator value changes) [8-9].
- Runtime information of the NS instance and each constituent VNF instance, accessible from the NS Info and each VNF Info.
- The entire set of NS-ILs and VNF-ILs available for use in the NSD and VNFD(s). These levels, built by NSD/VNFD developers at design time, cannot be changed at operation time. In case they need to be updated, DevOps strategies like those proposed in [15] could be used.
- Resource capacity information from each accessible VIM. This information can be found in the data repositories (see Fig. 2).

The DRPA applies the Auto Scaling Rules to the incoming performance/fault data. If they are not satisfied, NS scaling is required. In that case, the DRPA determines the NS-ILs that are candidate to satisfy the performance/fault criteria specified in the Auto Scaling Rules. Over these candidates, the DRPA applies the pertinent optimization criteria (e.g., minimize resource costs, energy consumption) and a set of constraints (e.g., available resource capacity, placement constraints) to output:
- The optimum NS-IL.
- The NFVI-PoPs that will accommodate the virtual resources associated to this optimum NS-IL. Moving from the current NS-IL towards the optimum NS-IL may entail the allocation and release of resources. For each new resource to be allocated, the DRPA selects the NFVI-PoP where this resource will be accommodated. Next, NFVO determines which VIM(s) provide access to the selected NFVI-PoPs.

Using these outputs, the NFVO triggers one of the following NS scaling procedures [11]:

- *VNF scaling*: One or more VNF instances in the NS instance modify their capacity by changing their VNF-ILs, and hence by adding and/or removing VNFC instances. This procedure assumes the new VNF-ILs are selected from the VNF flavor currently used.
- *Adding/Removing VNF instances*: Instances of existing VNFs are added/removed in the NS instance. For each VNF instance to be added, it is required to select its VNF-IL from a given VNF flavor.

As seen above, the specific procedure to be triggered is subjected to the differences that exist, in terms of VNF-ILs, between the two NS-ILs: the NS-IL of the NS instance, and the optimum NS-IL that DRPA has chosen. The possibility of choosing between different candidate NS-ILs (and hence triggering one of the above procedures) in the scaling operation adds flexibility compared to the autoscaling strategies present in the existing MANO solutions. The data models (TOSCA, YANG, or Heat Orchestration Templates [HOT]) used for their descriptors of NSs (and their constituents) have predefined, rigid autoscaling policies that do not allow choosing between different level of resources; instead, they set the level towards the NS (or one of its constituents) shall be scaled to.

## 4.2 Scaling operation workflow

As seen earlier, the goals of the VNF Scaling and Adding/Removing VNF instances procedures are different. However, the ways of performing them with MANO are very similar. Indeed, the addition/removal of VNF instances is no more than an extension of a VNF Scaling, with the peculiarity that the former implies (a) adding/removing all the VNFC instances of each VNF instance, and (b) instantiate/modify/remove VLs for VNF connectivity. Due to limited space, we concentrate on VNF Scaling in this paper.

In Fig. 5, we show the workflow messages for scaling a single VNF instance of an NS. These messages have been grouped into distinct phases: information collection, scaling triggering, resource allocation and resource release. Although the last two phases may be performed independently, it could happen both are required in the same scaling scenario (e.g., replacing a running VNFC instance by other instance of greater capacity). In that case, allocation goes before release to guarantee service continuity (e.g., when starting the new VNFC instance, the old instance can be deleted).

In the information collection phase, the NFVO gathers performance/fault data from VNFMs and VIMs. Performance metrics at NS/VNF level are reported with Performance Information Available Notifications, and performance metric-based threshold crossed values with Threshold Crossed Notification (steps 1-2). VNF Indicator values are changed by EMs, and notified to NFVO by VNFMs (step 3).

In the scaling triggering phase (step 4) the DRPA uses the above information, along with the information the NFVO has accessible from data repositories (NSD/VNFDs, NS/VNF Info, and NFVI resources state) to decide the optimum NS-IL. If we assume the optimum NS-IL differs from the existing one in the VNF-IL of a single VNF instance, a VNF scaling procedure will be triggered. This would happen, for example, if the DRPA decides to scale an instance of the NS shown in Fig. 1, moving it from NS-IL#1 to NS-IL#3 (see Fig. 4). This means scaling the VNF-B instance from VNF-IL#1 to VNF-IL#3. In other words, incrementing the capacity of the VNFC-B1 instance, while leaving the VNFC-B2 instance unmodified. Note that this change involves both resource allocation and resource release phases.

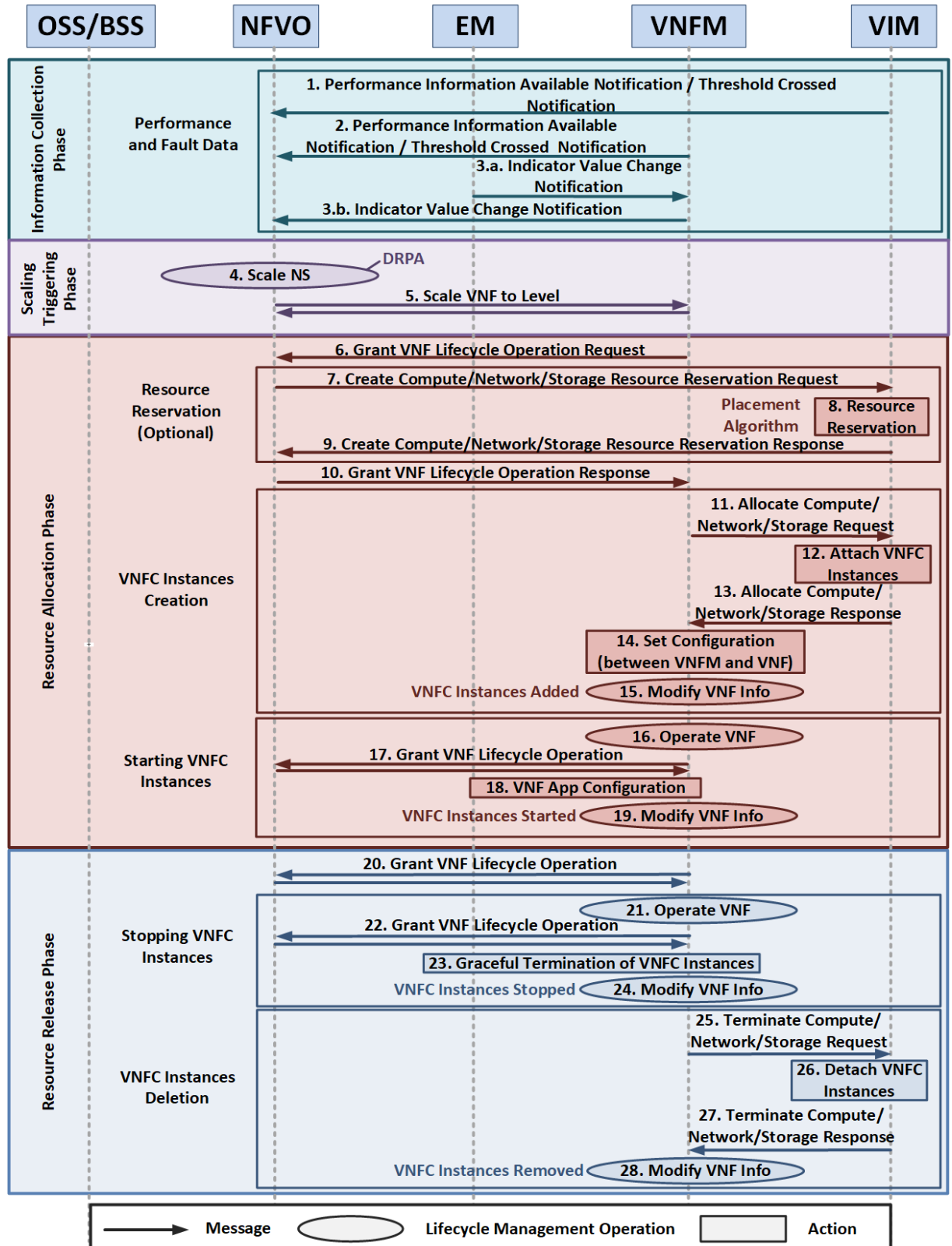

Figure 5. Workflow for the VNF Scaling Procedure.

The NFVO requires the VNFM to scale the VNF instance, sending it the new VNF-IL in the Scale VNF to Level Request[1]. Now, VNFM can initiate the scaling operation. To that end, the VNFM provides this lifecycle management operation with an unique ID using the Scale VNF to Level Response (step 5). The VNFM will use this ID to notify the NFVO the start and later the result of this operation[2]. Finally, the VNFM consults the VNF Info and the VNFD to compare the current VNF-IL against the new one. From this comparison, the resources to be allocated and/or released for this scaling operation can be derived.

In the resource allocation phase, we distinguish the following sub-phases:
- *Resource Reservation* (optional [12]): Prior to this sub-phase, VNFM asks NFVO for permission to allocate resources. To that end, VNFM sends NFVO the IDs of VDUs and internal VLs that map to the resources to be allocated (step 6). Although the NFVO already had this information after step 5, ETSI specifications impose this information exchange [9]. Then, the resource reservation sub-phase begins. From the output of the DRPA, the NFVO knows which NFVI-PoPs shall accommodate the resources to be allocated, and the VIM(s) providing access to those NFVI-PoPs. Now, these resources can be reserved for later allocation. Each selected VIM receives three reservation requests (step 7), one for each resource type (compute, storage, and network) it has to reserve in the NFVI-PoPs under its management. These requests include the placement constraints applicable to the specified resources. The VIM uses these constraints to perform a placement algorithm (step 8), deciding the appropriate NFVI-PoPs' resource zones [7] where resources are reserved. Then, the VIM sends the NFVO (step 9) the IDs of reserved resources. Finally, the NFVO sends the VNFM those IDs, and connectivity information [9] for each selected VIM. Now, VNFM knows how to access those VIMs, and which one may allocate each resource. If this sub-phase is not performed, two issues need to be considered. First, only steps 6 and 10 are executed. Secondly, the placement algorithm is now performed after resource allocation request (see next sub-phase).
- *VNFC Instances Creation*: The VNFM sends the reservations IDs to the corresponding VIMs (step 11) for resource allocation (step 12). At this point, VNFC instances have been created and their connectivity enabled. The IDs of the allocated resources are then sent to the VNFM (step 13). In step 14, the VNFM triggers the configuration of VNFC instances. Finally, VNFM updates the VNF Info in the data repository (step 15) to reflect the creation of new VNFC instances, and set their state to STOPPED.
- *Starting VNFC Instances*: To start the functionality of the new instances, the VNFM triggers an Operate VNF lifecycle operation (step 16). This operation will force (at the end of this sub-phase) the change of the instances' state from STOPPED to STARTED. Once the NFVO grants this operation (step 17), the new VNFC instances are configured at application level (step 18). To that end, VNFM communicates with the EM. Lastly, VNFM updates the VNF Info (step 19), changing the state of the new instances from STOPPED to STARTED. Note that some of the running VNFC instances could be affected by the creation of the new ones, and hence need to be (re)configured in terms of connectivity (e.g. new interfaces, updated link requirements) and/or application (e.g. sending/receiving packets to/from the new instances). In that case, the VNFM would order the corresponding VIM(s) and/or EM, respectively to make the necessary changes.

---

[1] Scale VNF operation message might also be used, but it has some limitations. See [9] for more information.

[2] Some lifecycle management operations require sending NFVO start and result notifications. For simplicity, all these notifications are omitted in the workflow.

For the Resource Release Phase, we have two sub-phases:
- *Stopping VNFC Instances*: In step 20, VNFM asks NFVO for permission to release resources. Then, VNFM triggers an Operate VNF lifecycle operation (step 21) to gracefully terminate some VNFC instances (forcing the stopping of VNFC instance at the end of this sub-phase). In step 23, affected instances are (re)configured (following counterpart strategies to those specified in the *Starting VNFC instances* sub-phase), and instances to be terminated are shut down. Finally, the state of stopped instances is changed from STARTED to STOPPED (step 24).
- *VNFC Instances Deletion*: The VNFM sends to the corresponding VIMs the IDs of the resources that host and connect the stopped VNFC instances (step 25). At this point, these instances are deleted (step 26). Then, the VIMs send back to the VNFM the resource IDs of released resources (step 27). After receiving those IDs, the VNFM updates the VNF Info (step 28) to reflect the instance deletion.

# 5. Conclusions

In this article, we shed light on the NS scaling operation with NFV. The options for automatically scaling a NS with the NFV framework are limited by the way the NSD is constructed. During its lifecycle, an NS instance only can move among the instantiation levels defined in the NSD, so their design is critical to ensure an effective automated scaling. In this work, we have analyzed how these levels are built in a NSD. To facilitate their understanding, we have proposed an NSD example, where different instantiation levels are included for scaling a NS.

We also have shown the different procedures the NFVO may trigger to scale an NS instance according to ETSI specifications, and how NFVO may automate them. To that end, the NFVO runs a DRPA that, taking NSD content and information of the operative NS instance, determines the optimum instantiation level towards the NS instance must be scaled to. This output forces the way the scaling procedures are performed in NFV. For one representative procedure, we have proposed an ETSI-compliant workflow that clarifies the interactions and information exchanges between the functional blocks in the NFV framework.

## Acknowledgments

This work is partially supported by the Spanish Ministry of Economy and Competitiveness, the European Regional Development Fund (Project TEC2016-76795-C6-4-R), the Spanish Ministry of Education, Culture and Sport (FPU Grant 16/03354), and the University of Granada, Andalusian Regional Government and European Social Fund under Youth Employment Program.

# Biographies

**Oscar Adamuz-Hinojosa** (oadamuz@ugr.es) received his B.Sc. and M.Sc. in Telecommunications Engineering from the University of Granada (Spain) in 2015 and 2017, respectively. He is currently working toward the Ph.D. degree with the Department of Signal Theory, Telematics and Communication of the University of Granada. His research interests are focused on SDN, NFV and network slicing in 5G systems.

**Jose Ordonez-Lucena** (jordonez@ugr.es) received his M.Sc. in Telecommunications Engineering in 2017 from the University of Granada, Spain. He is currently a Ph.D. student in the Department of Signal Theory, Telematics and Communications of the University of Granada. His research interests include network softwarization technologies (SDN/NFV), network slicing, and 5G mobile network architectures.

**Pablo Ameigeiras** (pameigeiras@ugr.es) received his M.Sc.E.E. degree in 1999 from the University of Malaga, Spain. He performed his master thesis at the Chair of Communication Networks, Aachen University (Germany). In 2000 he joined Aalborg University (Denmark), where he carried out his Ph.D. thesis. In 2006 he joined the University of Granada, where he has been leading several projects in the field of LTE and LTE Advanced systems. Currently his research interests include the application of the SDN/NFV paradigms for 5G systems.

**Juan J. Ramos-Munoz** (jjramos@ugr.es) received in 2001 his M.Sc. in Computer Sciences degree (2001) and Doctorate degree (2009) from the University of Granada, Spain. He is a Lecturer at the Department of Signals Theory, Telematics and Communications of the UGR. He is also member of the Wireless and Multimedia Networking Lab. His research interests are focused on real-time multimedia streaming, Quality of Experience assessment, network virtualization and network slicing for 5G.

**Diego Lopez** (diego.r.lopez@telefonica.com) joined Telefonica I+D in 2011 as a Senior Technology Expert and is currently in charge of the Technology Exploration activities within the GCTO Unit. Diego is focused on network virtualization, infrastructural services, network management, new network architectures, and network security. Diego chairs the ETSI ISG on Network Function Virtualization and the NFVRG within the IRTF, and he is member of the Board of 5TONIC, the Telefonica 5G Testbed.

**Jesus Folgueira** (jesus.folgueira@telefonica.com) received his M.Sc. degree in Telecommunications Engineering from UPM (1994) and MSc in Telecommunication Economics in 2015 (UNED). He joined Telefónica I+D in 1997. He is currently the Head of Transport and IP Networks within Telefonica Global CTO unit, in charge of Network Planning and Technology. He is focused on Optical, Metro & IP Networks architecture and technology, network virtualization and advanced switching. His expertise includes Broadband Access, R&D Management, and network deployment.